\documentclass[11pt]{article}
\pdfoutput=1

\usepackage{acl2016}
\usepackage{times}
\usepackage{amssymb}
\usepackage{amsfonts}
\usepackage{amsmath}
\usepackage{graphicx}
\usepackage{xspace}
\usepackage{xcolor}
\usepackage{pifont}
\usepackage{enumitem}
\usepackage{url}
\usepackage{latexsym}
\usepackage{array}

\aclfinalcopy 
\def\aclpaperid{1} 

\newcolumntype{C}[1]{>{\centering\let\newline\\\arraybackslash\hspace{0pt}}m{#1}}
\newcommand{\BioASQ}{\textsc{bioasq}\xspace}

\newcommand{\PubMed}{\textsc{pubmed}\xspace}
\newcommand{\QA}{\textsc{qa}\xspace}

\newcommand{\WordVec}{\textsc{word2vec}\xspace}
\newcommand{\Cent}{\textsc{c}ent\xspace}
\newcommand{\CentIDF}{\textsc{c}ent\textsc{idf}\xspace}
\newcommand{\CentIDFRWMDQ}{\textsc{c}ent\textsc{idf-rwmd-q}\xspace}
\newcommand{\CentIDFRWMDD}{\textsc{c}ent\textsc{idf-rwmd-d}\xspace}

\newcommand{\PubMedSE}{\textsc{p}ub\textsc{m}ed\textsc{se}\xspace}
\newcommand{\PubMedSERWMDQ}{\textsc{p}ub\textsc{m}ed\textsc{se-rwmd-q}\xspace}
\newcommand{\Hybrid}{\textsc{h}ybrid\xspace}

\newcommand{\RWMDQ}{\textsc{rwmd-q}\xspace}
\newcommand{\RWMDD}{\textsc{rwmd-d}\xspace}
\newcommand{\RWMDMAX}{\textsc{rwmd-max}\xspace}
\newcommand{\WMD}{\textsc{wmd}\xspace}
\newcommand{\UMLS}{\textsc{umls}\xspace}
\newcommand{\ANNCentIDFRWMDQ}{\textsc{ann-c}ent\textsc{idf-rwmd-q}\xspace}

\newcommand{\ANN}{\textsc{ann}\xspace}

\newcommand{\idf}{\textsc{idf}\xspace}
\newcommand{\MIP}{\textsc{mip}\xspace}
\newcommand{\MAP}{\textsc{map}\xspace}
\newcommand{\MAIP}{\textsc{maip}\xspace}
\newcommand{\nDCG}{n\textsc{dcg}\xspace}
\newcommand{\nDCGAtK}{n\textsc{dcg}@$k$\xspace}
\newcommand{\nDCGAtHundred}{n\textsc{dcg}@$100$\xspace}

\title{Using Centroids of Word Embeddings and Word Mover's Distance for Biomedical Document Retrieval in Question Answering}

\author{Georgios-Ioannis Brokos$^1$, Prodromos Malakasiotis$^{1,2}$ \and Ion Androutsopoulos$^{1,2}$\\ 
		$^1$Department of Informatics, Athens University of Economics and Business, Greece\\
		Patission 76, GR-104 34 Athens, Greece\\
		\url{http://nlp.cs.aueb.gr}\\
		$^2$Institute for Language and Speech Processing, Research Center `Athena', Greece\\
		Artemidos 6 \& Epidavrou, GR-151 25 Maroussi, Athens, Greece\\
		\url{http://www.ilsp.gr}
}

\date{}
\begin{document}
\maketitle

\begin{abstract}
We propose a 
document retrieval method for question answering  that represents documents and
questions as weighted centroids of word embeddings and reranks the retrieved documents with a relaxation of Word Mover's Distance. 
Using biomedical questions and documents from \BioASQ, we show that our method is competitive with \PubMed.
With a top-$k$ approximation, our method is fast, and easily portable to other domains and languages.  
\end{abstract}


\section{Introduction} \label{sec:introduction}

Biomedical experts (e.g., researchers, clinical doctors) routinely need to search the biomedical literature to support research hypotheses, treat rare syndromes, follow best practices etc. The most widely used biomedical search engine is \PubMed, with more than 24 million biomedical references and abstracts, mostly of journal articles.\footnote{See \url{http://www.ncbi.nlm.nih.gov/pubmed}.}
To improve their performance, biomedical search engines often use large, manually curated ontologies, e.g., to identify biomedical terms and expand queries with related terms.\footnote{\PubMed uses 
\UMLS 
(\url{http://www.nlm.nih.gov/research/umls/}). See also the  GoPubMed search engine (\url{http://www.gopubmed.com/}).} Biomedical experts, however, report that search engines often miss relevant documents and return many irrelevant ones.\footnote{Malakasiotis et al.\ \shortcite{Malakasiotis2015} summarize the findings of interviews that investigated how biomedical experts search.} 


There is also growing interest for biomedical question answering (\QA) systems \cite{Athenikos2010,Bauer2012,Tsatsaronis2015}, which allow their users to specify their information needs more precisely, as natural language questions
rather than Boolean queries, 
and aim to produce more concise answers.
Document retrieval is particularly important in biomedical \QA, since most of the information sought 
resides in documents and is essential in
later 
stages.

We propose a new document retrieval method. Instead of representing documents and 
questions as bags of words, we represent them as the centroids of their word embeddings \cite{Mikolov2013b,Pennington2014} and retrieve the documents whose centroids are closer to the centroid of the 
question. This allows retrieving relevant documents that may have no common terms with the 
question without query expansion. Using biomedical questions from the \BioASQ competition \cite{Tsatsaronis2015}, we show that our method combined with a relaxation of the recently proposed Word Mover's Distance (\WMD) \cite{Kusner2015} is competitive with \PubMed. We also show that with a top-$k$ approximation, our method is particularly fast, with no significant decrease in effectiveness. Given that it does not require ontologies, term extractors, or manually labeled training data, our method could be easily ported to other domains (e.g., legal texts) and languages.  


\section{The proposed method} \label{sec:data}

\begin{figure}
\centering
\includegraphics[height=4.87cm]{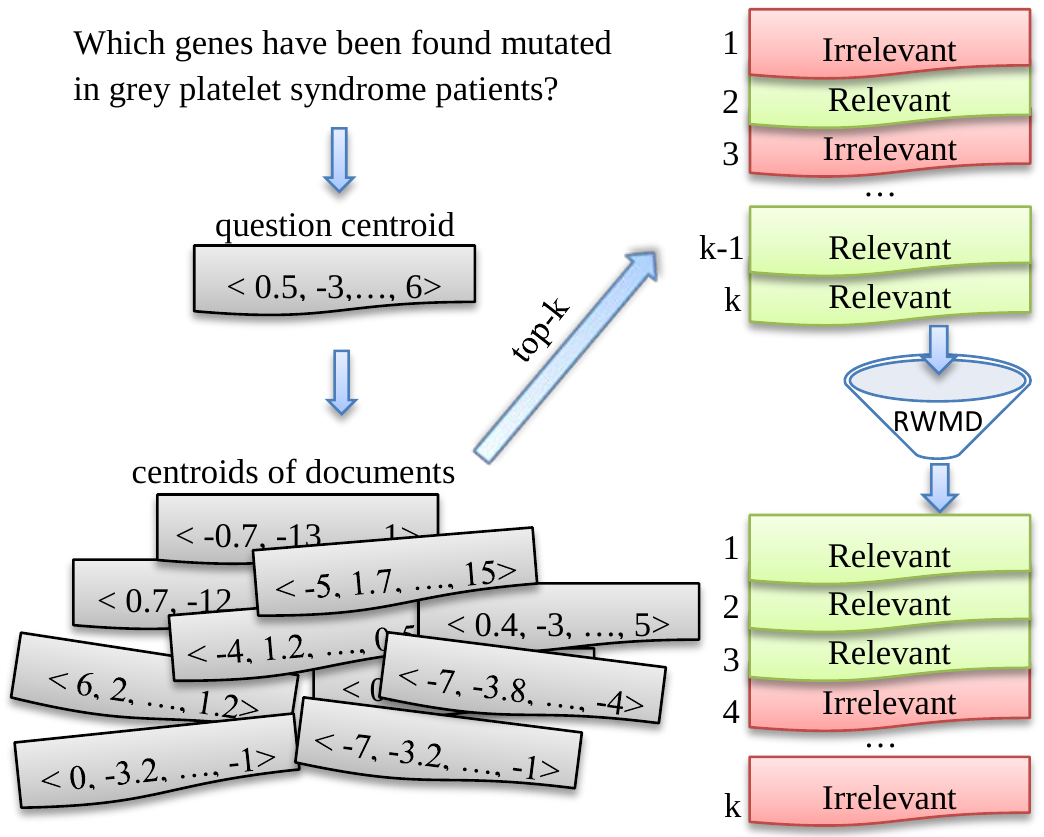}
\caption{Illustration of the proposed method.}
\label{fig:method_flow}
\end{figure}

The word embeddings and document centroids are pre-computed. For each 
question, its centroid is computed and the documents with the top-$k$ nearest (in terms of cosine similarity) centroids are retrieved (Fig.~\ref{fig:method_flow}). The retrieved documents are then optionally reranked using a relaxation of \WMD.

\subsection{Centroids of documents and 
questions}

In the simplest case, the centroid $\vec{t}$ of a text 
$t$ is the sum of the embeddings of the tokens of $t$ 
divided by the number of tokens in $t$. Previous work on hierarchical biomedical document classification \cite{Kosmopoulos2015} reported improved performance when the \idf scores of the tokens are also taken into account as follows: 
\begin{equation}
\vec{t} =  
\frac{\sum\limits_{j=1}^{|V|} \vec{w}_j \cdot \textrm{TF}(w_j, t) \cdot \textrm{IDF}(w_j)}
       {\sum\limits_{j=1}^{|V|} \textrm{TF}(w_j, t) \cdot \textrm{IDF}(w_j)}
\label{eq:idfCentroid}
\end{equation}
where $|V|$ is the vocabulary size (approx.~1.7 million words,
ignoring stop words),
$w_j$ is the $j$-th vocabulary word,
$\vec{w}_j$ its embedding,
 $\textrm{TF}(w_j, t)$ the term frequency 
of $w_j$ in $t$, and $\textrm{IDF}(w_j)$ the inverse document frequency of $w_j$ \cite{Manning2008}.
We use the 200-dimensional word embeddings 
of \BioASQ, obtained by applying \WordVec \cite{Mikolov2013b} to 
approx.~11 million abstracts from PubMed.\footnote{
The skipgram model of \WordVec was used, with hierarchical softmax, 5-word windows, and default other parameters. See \url{http://participants-area.bioasq.org/info/BioASQword2vec/} for further details.
} 
The \idf scores are computed on the 11 million abstracts. 


\subsection{Document retrieval and reranking} \label{sec:method}

Given a 
question with centroid $\vec{q}$, identifying the documents with the $k$ nearest centroids requires computing the distance between $\vec{q}$ and each document centroid, which is impractical for large document collections. Efficient approximate top-$k$ algorithms, however, 
exist. They divide the vector space into subspaces and use trees to index the instances in each subspace \cite{Arya1998,Indyk1998,Andoni2006,Muja2009}. We show that with an approximate top-$k$ algorithm, document retrieval is very fast, with no significant decrease in performance. The top-$k$ retrieved documents $d_i$ are ranked by decreasing (cosine) similarity of their centroids to $\vec{q}$. We call this method \Cent when the simple (no \idf) centroids are used, and \CentIDF when the \idf-weighted centroids (Eq.~\ref{eq:idfCentroid}) are used. 

The top-$k$ documents are optionally reranked with an approximation of the \WMD distance. \WMD measures the total distance the word embeddings of two texts (in our case, 
question and document) have to travel to become identical. In its full form, \WMD allows each word embedding to be partially aligned (travel) to multiple word embeddings of the other text, which requires solving a linear program and is too slow for our purposes. Kusner et al.\ \shortcite{Kusner2015} reported promising results in text classification using \WMD as the distance of a $k$-\textsc{nn} classifier. They also introduced relaxed, much faster 
\WMD versions. In our case, the first relaxation (\RWMDQ) sums the distances the word embeddings $\vec{w}$ of the 
question $q$ have to travel to the closest word embeddings $\vec{w}'$ of the document $d$:
\begin{equation}
\RWMDQ(q, d) = \sum_{w \in q} \min_{w' \in d} \mathit{dist}(\vec{w}, \vec{w}')
\label{eq:rwmdq}
\end{equation} 
Following Kusner et al., we use the Euclidean distance as $\mathit{dist}(\vec{w}, \vec{w}')$. Similarly, the second relaxed form (\RWMDD) sums the distances of the word embeddings of $d$ to the closest embeddings of $q$. If we set  $\mathit{dist}(\vec{w}, \vec{w}') = 1$ if $w, w'$ are identical and $0$ otherwise, \RWMDQ counts how many words of $q$ are present in $d$, and \RWMDD counts the words of $d$ that are present in $q$. Kusner et al.\ found the maximum of \RWMDQ and \RWMDD (\RWMDMAX) to be the best relaxation of \WMD. In our case, where $q$ is much shorter than $d$, \RWMDQ works much better, because $d$ contains many irrelevant words that have no close counter-parts in $q$, and their long distances dominate in \RWMDD and \RWMDMAX.\footnote{We do not report results with \RWMDMAX reranking, because they are as bad as results with \RWMDD.} We call \CentIDFRWMDQ and  \CentIDFRWMDD the \CentIDF method with the additional reranking by \RWMDQ or \RWMDD, respectively.


\section{Experiments}

\subsection{Data}

We used the 1,307 
training questions
and 
the gold relevant \PubMed document ids of the 
fourth year of \BioASQ (Task 4b).\footnote{The questions and gold document ids are available from \url{http://participants-area.bioasq.org/}. 
The 1,307 questions are all the training and test questions of the previous years of \BioASQ, which were available to the participants of the fourth year. We use all the 1,307 questions for testing, since our method is unsupervised.}
The questions were written by biomedical experts, who also identified the gold relevant documents using \PubMed, and reflect real 
needs \cite{Tsatsaronis2015}. 
We pass each question to our methods (after tokenization and stop-word removal) or the \PubMed search engine 
(hereafter \PubMedSE), which performs its own tokenization and query expansion.\footnote{We use relevance ranking
(not recency) in \PubMedSE.}

The document collection that we search contains 
approx.~14 million article abstracts and titles from the November 2015 \PubMed dump, which was also used in the fourth year of \BioASQ.\footnote{The dump is available from 
\url{https://www.nlm.nih.gov/databases/license/license.html}. The 14 million articles do not include approx.\ 10 million articles for which only titles are provided. There are hardly any title-only gold relevant documents, and \PubMedSE very rarely returns title-only documents.} 
Our methods view each document as a concatenation of the title and abstract of an article.\footnote{It is unclear to us if \PubMed also searches the full texts of the articles, which may put our methods at a disadvantage.} The titles and abstracts have an average length of 
approx.~13 and 143 tokens, respectively. When comparing against \PubMedSE, we ignore documents returned by \PubMedSE that are not in the dump, but this is very rare and does not affect the results. 


\subsection{Experimental results}

\begin{figure}
\centering
\includegraphics[width=\columnwidth]{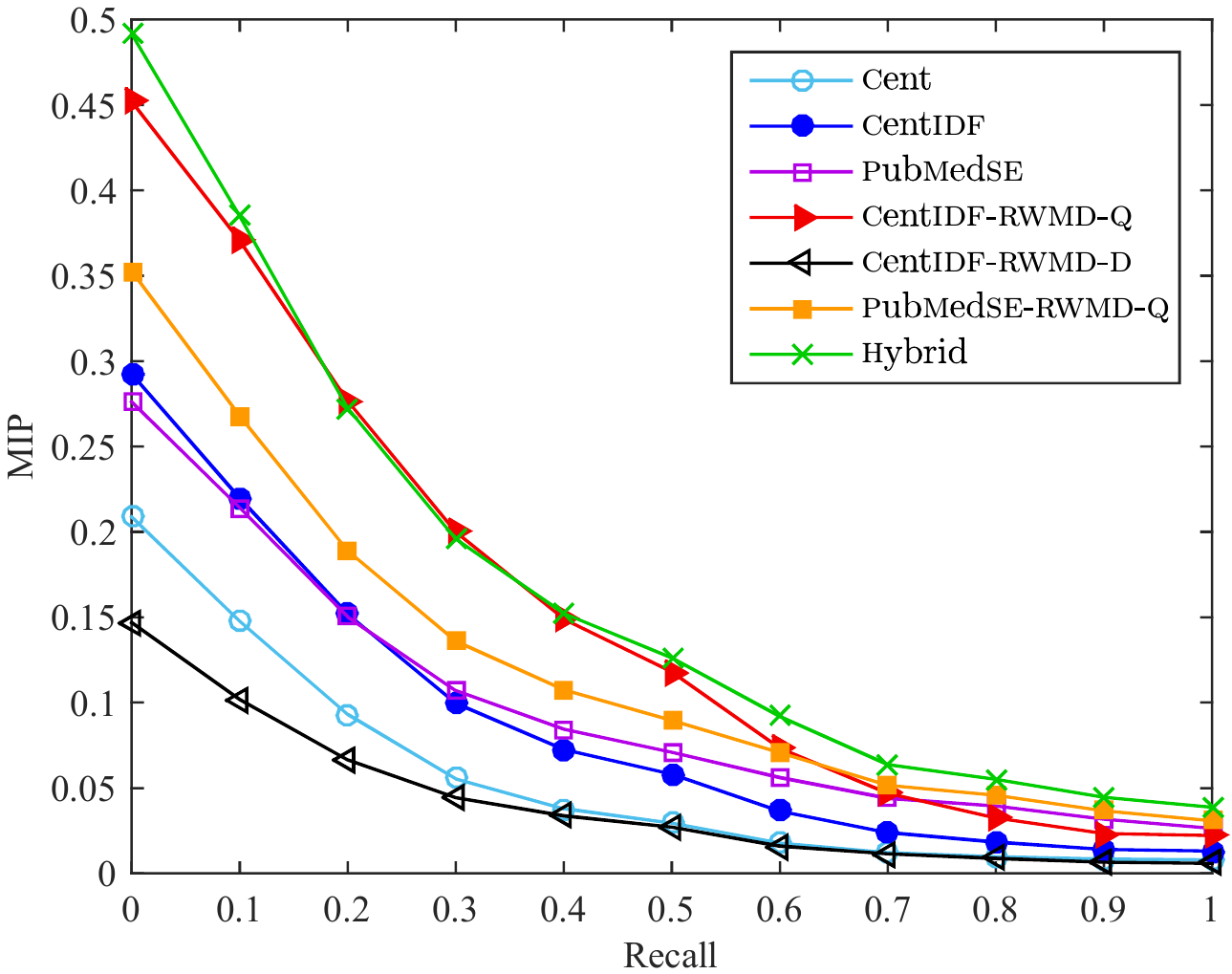}
\caption{Mean Interpolated Precision at 11 recall points, for $k$ (documents to retrieve) set to 1,000.}
\label{fig:mip@recall}
\end{figure}

\begin{figure}
\centering
\includegraphics[width=\columnwidth]{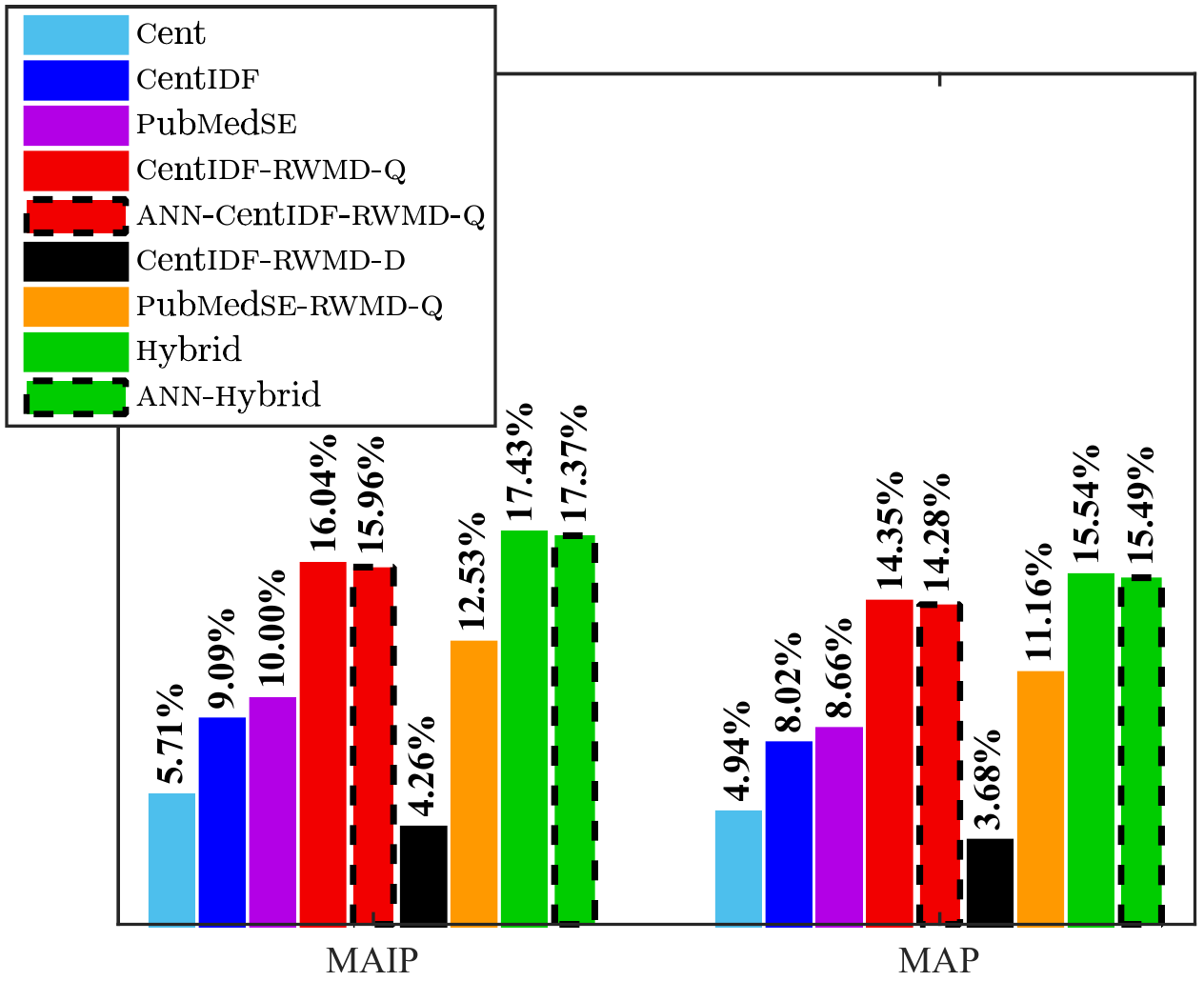}
\caption{\MAIP and \MAP scores, for $k$ (documents to retrieve) set to 1,000.}
\label{fig:maip_map}
\end{figure}

\begin{figure}
\centering
\includegraphics[height=6.4cm]{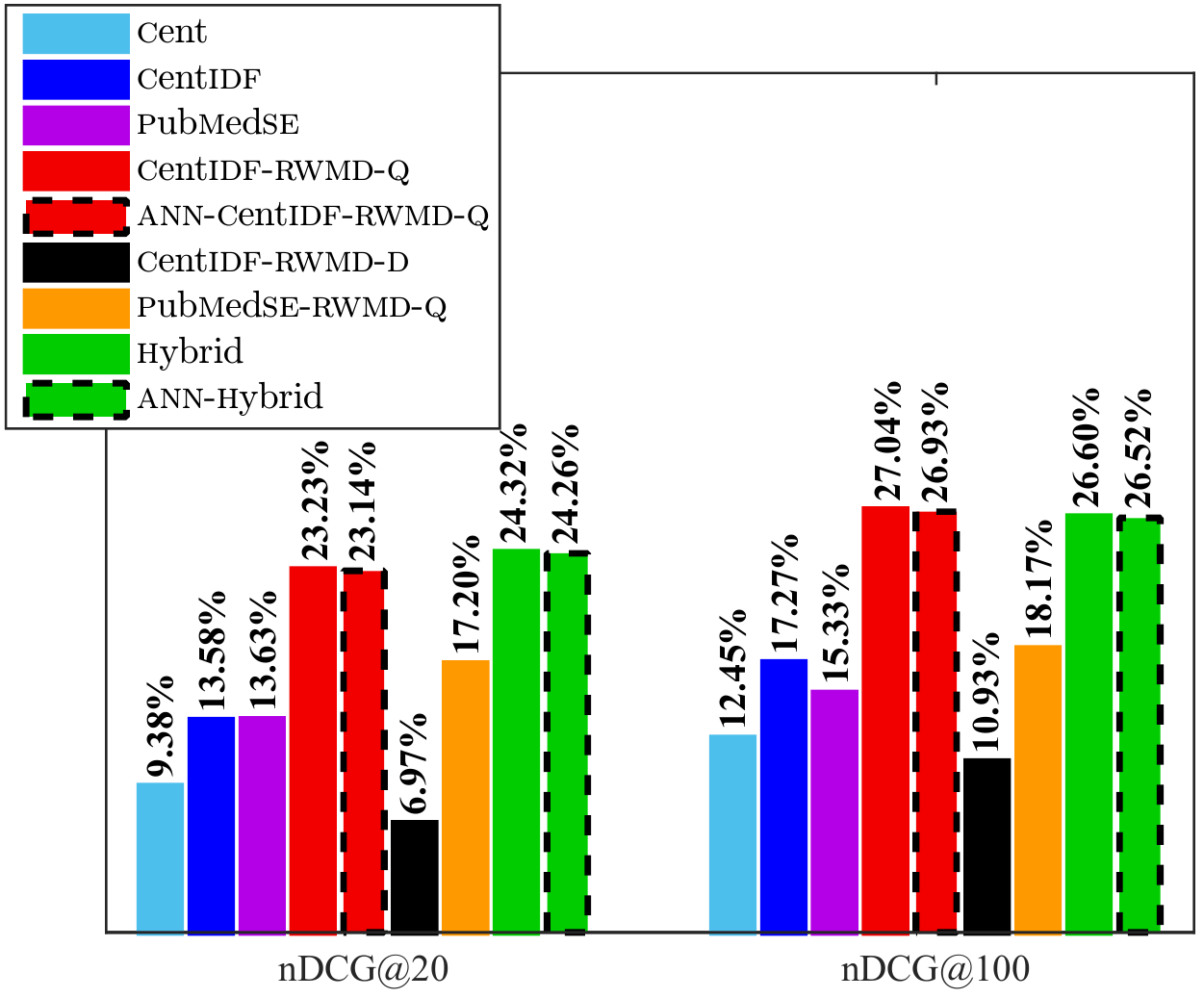}
\caption{\nDCGAtK, for $k$ = 20 and $k$ = 100.}
\label{fig:ndcg}
\end{figure}

Figures~\ref{fig:mip@recall}--\ref{fig:ndcg} show Mean Interpolated Precision (\MIP) at 11 recall levels, Mean Average Interpolated Precision (\MAIP),
Mean Average Precision (\MAP), and Normalized Discounted Cumulative Gain (\nDCG).\footnote{All measures are widely used \cite{Manning2008}. We use binary relevance in \nDCG, as in the \BioASQ dataset.} Roughly speaking, \MAIP 
is the area under the \MIP curve, 
\MAP is the same area without interpolation, 
and \nDCG is an alternative to \MAIP. Unless otherwise stated, the number of 
retrieved documents 
is set to $k = $ 1,000.

Figure~\ref{fig:mip@recall} shows that \Cent performs much worse than \CentIDF. At low recall, \CentIDF is as good as \PubMedSE, but \PubMedSE outperforms \CentIDF at high recall.
Reranking the top-$k$ documents of \CentIDF by \RWMDQ has a significant impact, leading to a system (\CentIDFRWMDQ) that performs better or as good as \PubMedSE up to 0.7 recall. Reranking the top-$k$ documents of \PubMedSE by \RWMDQ (\PubMedSERWMDQ) also improves the performance of \PubMedSE. Reranking the top-$k$ documents of \CentIDF by \RWMDD (or \RWMDMAX, not shown) leads to much worse results (\CentIDFRWMDD), for reasons already explained.\footnote{The same holds when the top-$k$ documents of \PubMedSE are reranked by \RWMDD or \RWMDMAX (not shown).} Similar conclusions are reached by examining the \MAIP, 
\MAP, and \nDCG scores.

Keyword-based information retrieval may miss relevant documents that use different terms than the 
question, even with query expansion. \PubMedSE retrieves no documents for 35\% (460/1307) of our questions.\footnote{The experts that identified the gold relevant documents used simple keyword, Boolean, and advanced \PubMedSE queries, whereas we used the English questions as queries.} Further experiments (not reported), however, indicate that \PubMedSE has higher precision than \CentIDFRWMDQ, when \PubMedSE returns documents, at the expense of lower recall. Hence, there is scope to combine \PubMedSE with our methods. As a first, crude step, we tested a method
(\Hybrid)
that returns the documents 
of \CentIDFRWMDQ when \PubMedSE retrieves no documents, and 
those of \PubMedSERWMDQ otherwise. \Hybrid had the best results in our experiments; the only exception was its
\nDCGAtHundred score, which was slightly lower than 
the score of  \CentIDFRWMDQ.

\begin{table}
\small
\centering
\begin{tabular}{|C{0.40\columnwidth}|C{0.13\columnwidth}|C{0.18\columnwidth}|C{0.09\columnwidth}|}
\hline System           & Search & Reranking & Total\\
\hline \CentIDFRWMDQ    & 47.41 $(\pm 1.22)$  & 14.45 $(\pm 6.15)$ & 61.86\\
\hline \ANNCentIDFRWMDQ & 0.36  $(\pm 0.04)$  & 14.24 $(\pm 6.06)$ & 14.60\\
\hline
\end{tabular}
\smallskip
\caption{Average times (in seconds) over all the questions of the dataset ($k$ = 1000).}
\label{tab:ann_times}
\end{table}

Table~\ref{tab:ann_times} shows 
that an approximate top-$k$ algorithm (\ANN) in \CentIDFRWMDQ (\ANNCentIDFRWMDQ) reduces dramatically the time to obtain the top-$k$ documents, with a very small decrease in \MAIP, \MAP, and \nDCG scores (Figures~\ref{fig:maip_map} and \ref{fig:ndcg}).\footnote{We use Annoy  (\url{https://github.com/spotify/annoy}), 100 trees, 1,000 neighbors, search-$k$ = $10 \cdot |\text{trees}| \cdot |\text{neighbors}|$. Times on a server with 4 Intel Xeon E5620 CPUs (16 cores total), at 2.4 GHz, with 128 GB RAM.}

We also compared against the other participants of the second year of \BioASQ; the participant results of later years are not yet available.\footnote{We used the evaluation platform of \BioASQ (\url{http://participants-area.bioasq.org/oracle}).} The official \BioASQ score is 
\MAP; \MIP, \MAIP, and \nDCG scores are not provided. Our best method was again \Hybrid (avg.\ \MAP over the five batches of the second year 16.18\%). It performed overall better than the \BioASQ `baselines' (best avg. \MAP 15.60\%) and all eight participants, except for the best one (avg.\ \MAP 28.20\%). The best system \cite{Choi2014} used dependency \textsc{ir} models \cite{Metzler2005}, combined with \UMLS and query expansion heuristics (e.g., adding the titles of the top-$k$ initially retrieved documents to the query). The `baselines' are actually very competitive; no system beat them in the first year, and only one was better in the second year. They are \PubMedSE, but using \BioASQ-specific heuristics (e.g., instructing \PubMedSE to ignore types of articles the experts did not consider). Our system is simpler and does not use
heuristics; hence, it can be ported more easily to other domains.

\section{Other related work}

Kosmopoulos et al.\ \shortcite{Kosmopoulos2015} reports that a $k$-\textsc{nn} classifier that represents articles as \idf-weighted centroids (Eq.~\ref{eq:idfCentroid}) of 200-dimensional word embeddings (200 features) is as good at assigning semantic labels (\textsc{m}e\textsc{sh} headings) to biomedical articles as when using millions of bag-of-word features, reducing significantly the training and classification times. To our knowledge, our work is the first attempt to use \idf-weighted centroids of word embeddings in information retrieval, and the first to use \WMD to rerank the retrieved documents. More elaborate methods to encode texts as vectors have been proposed \cite{Le2014,Kiros2015,Hill2016} and they could be used as alternatives to centroids of word embeddings, though the latter are simpler and faster to compute.

The \textsc{ohsumed} dataset \cite{Hersh1994} is often used in biomedical information retrieval experiments. It is much smaller (101 queries, approx.~350K documents) than the \BioASQ dataset that we used, but we 
plan to experiment with \textsc{ohsumed} in future work for completeness.


\section{Conclusions and future work}

We proposed a new 
\QA driven document retrieval method that represents documents and 
questions as \idf-weighted centroids of word embeddings. Combined with a relaxation of the \WMD distance, our method is competitive with \PubMed, without ontologies and query expansion. Combined with \PubMed, it performs better than \PubMed on its own. With a top-$k$ approximation, it is fast, and easily portable to other domains and languages.  

We plan to consider alternative dense vector encodings of documents and queries, textual entailment \cite{Bowman2015,Rocktaschel2016}, and full-text documents, where it may be necessary to extend \RWMDQ to take into account the proximity (density) of the words of the (now longer) document the query words are mapped to.



\section*{Acknowledgments} 

The work of the second author was funded by the Athens University of Economics and Business Research Support Program 2014-2015, ``Action 2: Support to Post-doctoral Researchers''.
  

\bibliography{references}

\begin{thebibliography}{}

\bibitem[\protect\citename{Andoni and Indyk}2006]{Andoni2006}
A.~Andoni and P.~Indyk.
\newblock 2006.
\newblock Near-optimal hashing algorithms for approximate nearest neighbor in
  high dimensions.
\newblock In {\em Proc.\ of the 47th Annual IEEE Symposium on Foundations of
  Computer Science}, pages 459--468, Washington, DC.

\bibitem[\protect\citename{Arya \bgroup et al.\egroup }1998]{Arya1998}
S.~Arya, D.~M. Mount, N.~S. Netanyahu, R.~Silverman, and A.~Y. Wu.
\newblock 1998.
\newblock An optimal algorithm for approximate nearest neighbor searching fixed
  dimensions.
\newblock {\em Journal of the ACM}, 45(6):891--923.

\bibitem[\protect\citename{Athenikos and Han}2010]{Athenikos2010}
S.~J. Athenikos and H.~Han.
\newblock 2010.
\newblock Biomedical question answering: A survey.
\newblock {\em Computer Methods and Programs in Biomedicine}, 99(1):1--24.

\bibitem[\protect\citename{Bauer and Berleant}2012]{Bauer2012}
M.~A. Bauer and D.~Berleant.
\newblock 2012.
\newblock Usability survey of biomedical question answering systems.
\newblock {\em Human Genomics}, 6(1)(17).

\bibitem[\protect\citename{Bowman \bgroup et al.\egroup }2015]{Bowman2015}
S.~R. Bowman, G.~Angeli, C.~Potts, and C.~D. Manning.
\newblock 2015.
\newblock A large annotated corpus for learning natural language inference.
\newblock In {\em Proc.\ of the 2015 Conference on Empirical Methods in Natural
  Language Processing}, Lisbon, Portugal.

\bibitem[\protect\citename{Choi and Choi}2014]{Choi2014}
S.~Choi and J.~Choi.
\newblock 2014.
\newblock Classification and retrieval of biomedical literatures: {SNUMedinfo}
  at {CLEF QA} track {BioASQ} 2014.
\newblock In {\em Proc.\ of the QA Lab of the 5th Conference and Labs of the
  Evaluation Forum}, pages 1283--1295, Valencia, Spain.

\bibitem[\protect\citename{Hersh \bgroup et al.\egroup }1994]{Hersh1994}
W.~Hersh, C.~Buckley, T.~J. Leone, and D.~Hickam.
\newblock 1994.
\newblock {OHSUMED}: An interactive retrieval evaluation and new large test
  collection for research.
\newblock In {\em Proc.\ of the 17th Annual International ACM SIGIR Conference
  on Research and Development in Information Retrieval}, pages 192--201,
  Dublin, Ireland.

\bibitem[\protect\citename{Hill \bgroup et al.\egroup }2016]{Hill2016}
F.~Hill, K.~Cho, and A.~Korhonen.
\newblock 2016.
\newblock Learning distributed representations of sentences from unlabelled
  data.
\newblock {\em arXiv preprint 1602.03483}.

\bibitem[\protect\citename{Indyk and Motwani}1998]{Indyk1998}
P.~Indyk and R.~Motwani.
\newblock 1998.
\newblock Approximate nearest neighbors: Towards removing the curse of
  dimensionality.
\newblock In {\em Proc.\ of the 30th Annual ACM Symposium on Theory of
  Computing}, pages 604--613, Dallas, TX.

\bibitem[\protect\citename{Kiros \bgroup et al.\egroup }2015]{Kiros2015}
R.~Kiros, Y.~Zhu, R.~R. Salakhutdinov, R.~Zemel, R.~Urtasun, A.~Torralba, and
  S.~Fidler.
\newblock 2015.
\newblock Skip-thought vectors.
\newblock In {\em Advances in Neural Information Processing Systems 28}, pages
  3276--3284. Montr\'{e}al, Canada.

\bibitem[\protect\citename{Kosmopoulos \bgroup et al.\egroup
  }2016]{Kosmopoulos2015}
A.~Kosmopoulos, I.~Androutsopoulos, and G.~Paliouras.
\newblock 2016.
\newblock Biomedical semantic indexing using dense word vectors in {BioASQ}.
\newblock {\em Journal Of Biomedical Semantics, Supplement On Biomedical
  Information Retrieval}.
\newblock To appear.

\bibitem[\protect\citename{Kusner \bgroup et al.\egroup }2015]{Kusner2015}
M.~Kusner, Y.~Sun, N.~Kolkin, and K.~Q. Weinberger.
\newblock 2015.
\newblock From word embeddings to document distances.
\newblock In {\em Proc.\ of the 32nd International Conference on Machine
  Learning}, pages 957--966, Lille, France.

\bibitem[\protect\citename{Le and Mikolov}2014]{Le2014}
Q.~Le and T.~Mikolov.
\newblock 2014.
\newblock Distributed representations of words and phrases.
\newblock In {\em Proc.\ of the 31st International Conference on Machine
  Learning}, pages 1188--1196, Beijing, China.

\bibitem[\protect\citename{Malakasiotis \bgroup et al.\egroup
  }2014]{Malakasiotis2015}
P.~Malakasiotis, I.~Androutsopoulos, A.~Bernadou, N.~Chatzidiakou, E.~Papaki,
  P.~Constantopoulos, I.~Pavlopoulos, A.~Krithara, Y.~Almyrantis,
  D.~Polychronopoulos, A.~Kosmopoulos, G.~Balikas, I.~Partalas, G.~Tsatsaronis,
  and N.~Heino.
\newblock 2014.
\newblock Challenge {E}valuation {R}eport 2 and {R}oadmap.
\newblock {BioASQ} deliverable D5.4.

\bibitem[\protect\citename{Manning \bgroup et al.\egroup }2008]{Manning2008}
C.D. Manning, P.~Raghavan, and H.~Sch{\"u}tze.
\newblock 2008.
\newblock {\em Introduction to Information Retrieval}.
\newblock Cambridge University Press.

\bibitem[\protect\citename{Metzler and Croft}2005]{Metzler2005}
D.~Metzler and W.B. Croft.
\newblock 2005.
\newblock A {M}arkov {R}andom {F}ield model for term dependencies.
\newblock In {\em Proc.\ of the 28th Annual International ACM SIGIR conference
  on Research and Development in Information Retrieval}, pages 472--479,
  Salvador, Brazil.

\bibitem[\protect\citename{Mikolov \bgroup et al.\egroup }2013]{Mikolov2013b}
T.~Mikolov, W.~Yih, and G.~Zweig.
\newblock 2013.
\newblock Distributed representations of words and phrases and their
  compositionality.
\newblock In {\em Proc.\ of the Conference on Neural Information Processing
  Systems}, Lake Tahoe, NV.

\bibitem[\protect\citename{Muja and Lowe}2009]{Muja2009}
M.~Muja and D.~G. Lowe.
\newblock 2009.
\newblock Fast approximate nearest neighbors with automatic algorithm
  configuration.
\newblock In {\em Proc.\ of the International Conference on Computer Vision
  Theory and Applications}, pages 331--340, Lisboa, Portugal.

\bibitem[\protect\citename{Pennington \bgroup et al.\egroup
  }2014]{Pennington2014}
J.~Pennington, R.~Socher, and C.~D. Manning.
\newblock 2014.
\newblock Glo{V}e: Global vectors for word representation.
\newblock In {\em Proc.\ of the Conference on Empirical Methods on Natural
  Language Processing}, Doha, Qatar.

\bibitem[\protect\citename{Rockt{\"a}schel \bgroup et al.\egroup
  }2016]{Rocktaschel2016}
T.~Rockt{\"a}schel, E.~Grefenstette, K.~M. Hermann, T.~Ko{\v{c}}isk{\`y}, and
  P.~Blunsom.
\newblock 2016.
\newblock Reasoning about entailment with neural attention.
\newblock In {\em International Conference on Learning Representations}, San
  Juan, Puerto Rico.

\bibitem[\protect\citename{Tsatsaronis \bgroup et al.\egroup
  }2015]{Tsatsaronis2015}
G.~Tsatsaronis, G.~Balikas, P.~Malakasiotis, I.~Partalas, M.~Zschunke, M.R.
  Alvers, D.~Weissenborn, A.~Krithara, S.~Petridis, D.~Polychronopoulos,
  Y.~Almirantis, J.~Pavlopoulos, N.~Baskiotis, P.~Gallinari, T.~Artieres,
  A.~Ngonga, N.~Heino, E.~Gaussier, L.~Barrio-Alvers, M.~Schroeder,
  I.~Androutsopoulos, and G.~Paliouras.
\newblock 2015.
\newblock An overview of the {BioASQ} large-scale biomedical semantic indexing
  and question answering competition.
\newblock {\em BMC Bioinformatics}, 16(138).

\end{thebibliography}
\bibliographystyle{acl2016}
\end{document}